\documentclass[onecolumn,10pt]{IEEEtran}

\usepackage{amsmath}
\usepackage{theorem}
\usepackage{amsmath}
\usepackage{mathrsfs}
\usepackage{epsfig}
\usepackage{amsfonts}
\usepackage{amssymb}
\usepackage{mathrsfs}
\usepackage{euscript}
\usepackage{graphicx}
\usepackage{multirow}
\usepackage{cite}

\begin{document}
\title{A Non-MDS Erasure Code Scheme For Storage Applications}
\author{
\authorblockN{Abbas Kiani, Soroush Akhlaghi}\\
\authorblockA{Shahed University, Tehran, Iran\\
Email: \{akiani,akhlaghi\}@shahed.ac.ir}}
\maketitle
\begin{abstract}
This paper investigates the use of redundancy and self repairing against node failures in distributed storage systems, using various strategies. In replication method, access to one replication node is sufficient to reconstruct a lost node, while in MDS erasure coded systems which are optimal in terms of redundancy-reliability tradeoff, a single node failure is repaired after recovering the entire stored data. Moreover, regenerating codes yield a tradeoff curve between storage capacity and repair bandwidth. The current paper aims at investigating a new storage code. Specifically, we propose a non-MDS $(2k,k)$ code that tolerates any three node failures and more importantly, it is shown using our code a single node failure can be repaired through access to only three nodes.
\end{abstract}

\section{Introduction}
The field of large scale data storage has witnessed significant growth in recent years with applications such as social networks and file sharing. In Storage systems, data should be stored over multiple nodes (independent storage devices such as disks, servers or peers) and it may happen a storage node is failed or leaves the system. Thus, a reliable storage capability over individually unreliable nodes can be achieved through introducing redundancy.

There are various strategies for distributing redundancy and depending on the used method the system can tolerate a limited number of node failures. Moreover, to keep the redundancy the same as if there is no node failures, the system should have self-repairing capability. In other words, each damaged node is replaced with a new node after transferring data over the network. Reconstructing a failed node and the maintenance bandwidth are called repair problem and repair bandwidth, respectively.

Erasure codes are the most common strategy for distributing redundancy. An erasure coded system employs totally $n$ packets of the same size, $k$ of which are data packets
(the fragments of the original data file) and $n-k$ of which are parity packets (the coding information). It is worth mentioning that the process of coding can be done using MDS or non-MDS codes. In a distributed storage system, these packets are stored at $n$ different nodes over the network. MDS codes~\cite{MDS} are optimally space-efficient and the encoding process is such that having access to any $k$ nodes is adequate to recover the original data file. In these codes, each parity node increases fault tolerance. In other words, a $(n,k)$ MDS coded system can tolerate any $n-k$ node failures.

Replication, RAID 5, RAID 6~\cite{RAID}, and Reed-Solomon codes \cite{ReedSolomon} are the most popular MDS codes that have been used in storage systems.
In replication, the parity nodes and data nodes are the same. In fact, each data node has a replica which is stored in a related parity node.
RAID 5 and RAID 6 employ $n-k=1$ and $n-k=2$ parity node respectively; however, Reed-Solomon codes can be designed for any value of $(n,k)$~\cite{ReedSolomon}.
Another class of MDS codes are MDS array codes such as EVENODD~\cite{evenodd}, extended EVENODD~\cite{extendedevenodd}, Row-Diagonal Parity (RDP)~\cite{RDP}, X-code~\cite{Xcode}, P-code~\cite{Pcode}, B-code~\cite{Bcode}, and STAR code~\cite{Starcode}. These codes are based on XOR operation and have lower encoding and decoding complexity than Reed-Solomon codes.
\begin{figure*}
\centering
\epsfig{file=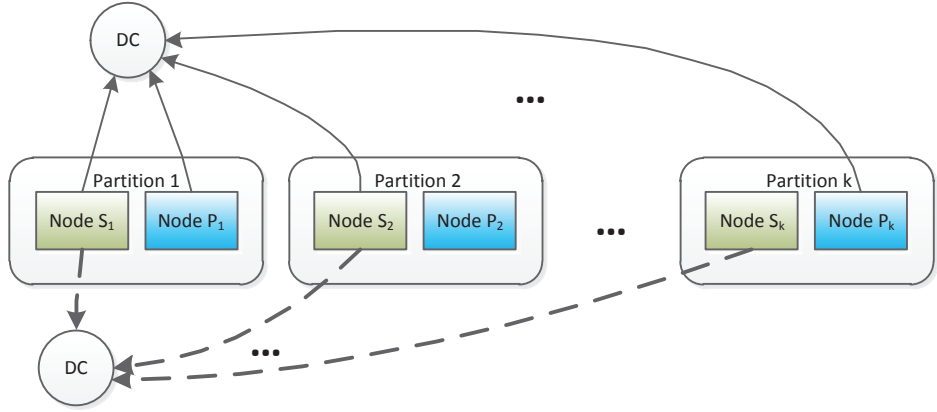,width=0.75\linewidth,clip=}
\caption{The graphical representation of the proposed code. The recovery of original data file can be achieved by connecting to: (i) two nodes from a partition and $k-2$ different nodes selected over $k-2$ different partitions (solid-lines) and (ii) $2m~\leq~k$ parity nodes and $k-2m$ systematic nodes selected from $k$ different partitions (dashed-lines are the specific case $m=0$).}
\label{fig:1}
\end{figure*}

In~\cite{LDPC}, Low Density Parity-Check (LDPC) codes as a class of non-MDS codes are introduced. These codes aim at reducing encoding and decoding costs
computation over lossy networks; however, are not as space-efficient as MDS codes. Non-MDS codes are further investigated in several papers. As a case in point, Hafner in~\cite{WEAVER} proposes a new class of non-MDS XOR-based codes, called WEAVER codes. The WEAVER codes are vertical codes which can tolerate up to 12 node failures. In a vertical code like X-code and WEAVER code each node contains both data and parity packets. In contrast, each node in a flat-XOR code such as EVENODD, holds either data or parity packets. The authors in~\cite{combination} describe construction of two novel flat XOR-based code, called stepped combination and HD-combination codes. Also in~\cite{combination}, chain codes, a variant of chained configuration method~\cite{chain}, are investigated.

The standard MDS codes in terms of repair problem are inefficient and recreating a failed node consumes a repair bandwidth equal the entire stored data. This motivated Dimakis et al. in~\cite{RC} to propose a repair optimal MDS code, called regenerating code, to make a tradeoff between repair bandwidth and storage capacity per node. It is shown in~\cite{RC} that any point on the identified tradeoff curve can be achieved through the use of network coding. Furthermore, in~\cite{GRC} an extension of regenerating codes, dubbed generalized regenerating codes, are introduced for the case of having different download cost associated with each node. Moreover, the authors in~\cite{SRC} investigate the case in which the newcomer node can wisely select the existing node to connect to.

The repair model presented in~\cite{RC} is a functional repair. In the functional repair model the recreated packets stored at replaced node can be different with the lost packets. Contrast the functional repair with the exact repair in which each lost packet is exactly reconstructed. The exact repair for the minimum bandwidth regenerating codes is investigated in~\cite{exactMBR}. Also in~\cite{exactMSR1},~\cite{exactMSR2},~\cite{exactMSR3}, the exact repair for the minimum storage regenerating codes is addressed based upon the interference alignment concepts.

Regenerating codes outperform existing MDS erasure codes in terms of maintenance bandwidth; however, the constructing a new packet requires communication with $d~\geq~k$ nodes and the minimum repair bandwidth can be achieved when $d=n-1$. In addition, the surviving nodes have to apply random linear network coding to their packets. Accordingly, many of the proposed constructions require a huge finite-field size and are not feasible for practical storage systems. The current study aims to introduce a $(n,k)=(2k,k)$  non-MDS XOR-based code which can tolerate any three node failures. Accordingly, it is shown in this code a single node failure can be repaired through access to only three nodes regardless of $k$.

The rest of paper is organized as follows: Section~\ref{sec:cons} states the construction and motivates the main idea. In section~\ref{sec:RP}, we explain the repair problem of the proposed code. Finally, sections~\ref{sec:conclusion}, concludes the paper.

\section{Construction}\label{sec:cons}
In this section we describe the construction of the proposed non-MDS code. Fig.~\ref{fig:1} shows a graphical representation for this code. This code is a class of flat XOR-codes which contains $2k$ storage nodes and each node stores one packet. The construction is such that $k$ out of $2k$ existing nodes i.e., $\{S_i\}_{i=1,\ldots,k}$, hold data fragments, called systematic nodes. The remaining $k$ nodes, i.e.,$\{P_i\}_{i=1,\ldots,k}$, are the parity nodes which store parity packets. Also it is assumed that each systematic node $(S_i)$ has a related parity node $(P_i)$ in which they stand in a same partition. Thus, with this construction, the code entails $k$ partition.

\begin{figure*}
\centering
\epsfig{file=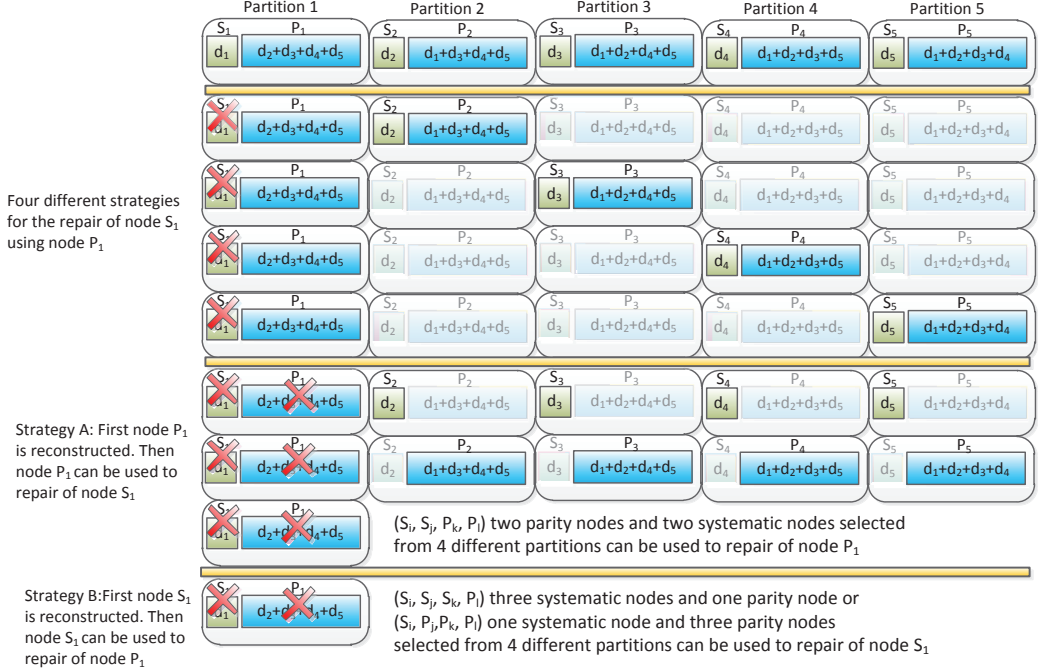,width=0.8\linewidth,clip=}
\caption{The repair problem of a $(n,k)=(10,5)$ code. The lost packet $d_1$ can be repaired by the use of three packets including its related parity packet
i.e, $d_2+d_3+d_4+d_5$. Also when $d_2+d_3+d_4+d_5$ has failed $d_1$
 can be reconstructed by the use of four nodes from another partitions.
}
\label{fig:2}
\end{figure*}

For storing a file of size $M$ using this construction, the file is divided to $k$ fragments i.e., $d_1,d_2,d_3,\ldots,d_k$, each of size $\frac{M}{k}$. Each fragment can be a single bit or a block of bits. These fragments are stored at $k$ systematic nodes. Fig.~\ref{fig:2} illustrates a $(n,k)=(10,5)$ code corresponding to the explained construction. Referring to Fig.~\ref{fig:2}, the five data fragments, i.e., $d_1,d_2,d_3,d_4$ and $d_5$, are stored at nodes $S_1,S_2,S_3,S_4$ and $S_5$ respectively. Noting the parity packet $p_i$ to be stored in parity node $P_i$ is computed as
\begin{equation}\label{equ1}
p_i=\sum_{j=1,~j\neq~i}^{k}d_i~,
\end{equation}
For $i=1,\ldots,k$. The addition here is bit-by-bit XOR for two data packets. For instance in a (10,5) code, as can be seen in Fig.~\ref{fig:2}, parity packets $p_1=d_2+d_3+d_4+d_5$, $p_2=d_1+d_3+d_4+d_5$, $p_3=d_1+d_2+d_4+d_5$, $p_4=d_1+d_2+d_3+d_5$ and $p_5=d_1+d_2+d_3+d_4$ are stored in parity nodes $P_1,P_2,P_3,P_4$ and $P_5$ respectively. It is worth mentioning that for specific case $k=2$ this code performs similar to replication method. Also for $k=3$ the parity packets are same with the parity packets of the proposed chain code in~\cite{combination}.
Now we are ready to discuss how the recovery of the original file can be achieved. It is assumed corresponding to a request to reconstructing
the original data file a Data Collector (DC) is initiated and connects to existing nodes. With this construction, DC requires to connect to at least $k$ out of existing nodes. Recall the proposed construction in this paper is non-MDS and having access to any $k$ nodes out of existing $2k$ nodes does not ensure restoring the original file. Each data collector has two possible strategies for selecting $k$ storage nodes to connect to: (i) DC can connect to both systematic node and parity node from a partition and $k-2$ different nodes selected from $k-2$ different partitions out of the $k-1$ remaining partitions (solid-lines in Fig.~\ref{fig:1} are a specific case of this scenario). When using this strategy there are
\begin{equation}\label{equ2}
\binom{k}{k-1}~\binom{k-1}{1}~2^{k-2}~=~(k)(k-1)2^{k-2}~,
\end{equation}
options for DC to choose $k$ nodes to connect to.
(ii) DC can connect to $2m~\leq~k$ parity nodes and $k-2m$ systematic nodes selected from $k$ different partitions. (dashed-lines in Fig.~\ref{fig:1} can be considered as a specific case of this scenario i.e. $m=0$). With strategy (ii), number of possible ways to choose $k$ nodes is computed as
\begin{equation}\label{equ3}
\sum_{m=0}^{\lfloor\frac{k}{2}\rfloor}~\binom{k}{2m}~=~2^{k-1}~.
\end{equation}
Thus, there totally exist $2^{k-2}(k^2-k+2)$ ways to recover the original file using $k$ nodes. Considering the two possible strategies, the proposed $(2k,k)$ code can tolerate any three node failures. Moreover, this code can tolerate up to $k-1$ node failures if these nodes are failed from $k-1$ different partitions.

As discussed, the storage per node for storing a file of size $M$ is $\frac{M}{k}$ which is equivalent with standard MDS codes and Minimum Storage Regenerating (MSR) codes.\footnote{The identified tradeoff curve in~\cite{RC} has two extremal points; one end of this curve corresponds to
the minimum storage per node and the other end corresponds to minimum bandwidth point. These two extremal points can
be achieved by the use of Minimum Storage Regenerating (MSR) and Minimum Bandwidth Regenerating (MBR) codes,
respectively.} However, for $2k\frac{M}{k}=2M$ total storage, MSR and standard MDS codes offer higher reliability. Recall to keep the reliability same across time, each failed node should be repaired. In the naive method that can be used to any MDS code, a single node repair can be done after transferring the whole data file over the network (the repair bandwidth is equal to $M$). Regenerating codes can reduce the repair bandwidth if we allow the new node connect to $d~>~k$ nodes. Our goal is reduce the repair bandwidth compared to the naive method when new node connects to $d~<~k$ nodes. The following section aims at addressing the repair model of suggested code.

\section{Repair problem}\label{sec:RP}
Note that when a node is failed or leaves the system a new node is initiated, attempting to connect to existing
nodes to reconstruct the failed node. During the course of repairing a damaged node, we face two scenarios: (i) The parity or systematic node which has common partition with the failed node (related node) is active and (ii) The related node has failed. In the case of existence of the related node, the failed node can be reconstructed by communicate to only three nodes i.e., the related node and both parity node and systematic node from another active partition. In fact, there are $k-1$ ways to repair a failed node through downloading from only three nodes. For example, referring to Fig.~\ref{fig:2}, we assume that the systematic node $S_1$ which holds data fragment $d_1$ is failed. When parity node $P_1$ which stores parity packet $d_2+d_3+d_4+d_5$ is active, the new node can restore $d_1$ through downloading three packets in $k-1=4$ ways as
\begin{eqnarray*}
(d_2+d_3+d_4+d_5)+(d_2)+(d_1+d_3+d_4+d_5)\nonumber\\
(d_2+d_3+d_4+d_5)+(d_3)+(d_1+d_2+d_4+d_5)\nonumber\\
(d_2+d_3+d_4+d_5)+(d_4)+(d_1+d_2+d_3+d_5)\nonumber\\
(d_2+d_3+d_4+d_5)+(d_5)+(d_1+d_2+d_3+d_4)
\end{eqnarray*}
Thus, in scenario (i), three nodes are involved during the course of downloading for reconstructing a new node. This leads to have a repair bandwidth equal to $3\frac{M}{k}$.

As discussed earlier, in MSR codes the new node should connect $d~\geq~k$ nodes to ensure reconstructing a failed node. In these codes the repair bandwidth is computed as $\frac{Md}{k(d-k+1)}$ which is a decreasing function with respect to $d$~\cite{RC} and, hence, when new node connects to the minimum possible nodes i.e., $k$ nodes, the repair bandwidth takes its maximum value i.e., $M$. For instance, in a $(n,k)=(10,5)$ MSR code, the repair bandwidth $\frac{3M}{5}$ can be achieved if new node connects to $d=6$ nodes which is greater than $d=3$ nodes in the proposed scheme.

For the case of scenario (ii), we can consider two different strategies. In the first strategy, dubbed strategy A, first the parity node is repaired and then used to reconstruct the related systematic node. For recreating the parity node without using the related systematic node, the new node should connect $2m$ parity nodes and $k-1-2m$ systematic nodes over $k-1$ different partition and there are
\begin{equation}\label{equ4}
\sum_{m=0}^{\lfloor\frac{k-1}{2}\rfloor}~\binom{k-1}{2m}~=~2^{k-2}~,
\end{equation}
ways to choose these nodes. For instance, referring Fig.~\ref{fig:2}, in a $(10,5)$ code there are $2^{(5-2)}=8$ ways in which the new node can use 0,2 or 4 parity nodes to repair parity node $P_1$ which stores $d_2+d_3+d_4+d_5$ without the use of node $S_1$. These eight ways are as
\begin{eqnarray*}
(d_2)+(d_3)+(d_4)+(d_5)~~~~~~~~~~~~~~~~~~~~~~~~~~~~~~~~~~~~~~~\nonumber\\
(d_1+d_3+d_4+d_5)+(d_1+d_2+d_4+d_5)+(d_4)+(d_5)\nonumber\\
(d_1+d_3+d_4+d_5)+(d_1+d_2+d_3+d_5)+(d_3)+(d_5)\nonumber\\
(d_1+d_3+d_4+d_5)+(d_1+d_2+d_3+d_4)+(d_3)+(d_4)\nonumber\\
(d_1+d_2+d_4+d_5)+(d_1+d_2+d_3+d_5)+(d_2)+(d_5)\nonumber\\
(d_1+d_2+d_4+d_5)+(d_1+d_2+d_3+d_4)+(d_2)+(d_4)\nonumber\\
(d_1+d_2+d_3+d_5)+(d_1+d_2+d_3+d_4)+(d_2)+(d_3)\nonumber\\
(d_1+d_3+d_4+d_5)+(d_1+d_2+d_4+d_5)+~~~~~~~~~~~~~~\nonumber\\
\!\!\!(d_1+d_2+d_3+d_5)+(d_1+d_2+d_3+d_4)~~~~~~~~~~~~~~~~~
\end{eqnarray*}
In the second strategy, called strategy B, first the failed systematic node is repaired and then involved in the reconstruction of the related parity node. When strategy B is employed, the new node requires to communicate with $2m+1$ parity nodes and $k-2m-2$ systematic nodes over $k-1$ different partitions and the number of possible ways to select these surviving nodes are computed as
\begin{equation}\label{equ5}
\sum_{m=1}^{\lceil\frac{k-1}{2}\rceil}~\binom{k-1}{2m-1}~=~2^{k-2}~.
\end{equation}
For example, as can be seen in Fig.~\ref{fig:2}, there exist $2^{(5-2)}=8$ options for the new node to choose one or three parity nodes for the repair of systematic node $S_1$ which holds $d_1$ when $P_1$ can not be involved. These options are as
\begin{eqnarray*}
(d_1+d_3+d_4+d_5)+(d_3)+(d_4)+(d_5)~~~~~~\nonumber\\
(d_1+d_2+d_4+d_5)+(d_2)+(d_4)+(d_5)~~~~~~\nonumber\\
(d_1+d_2+d_3+d_5)+(d_2)+(d_3)+(d_5)~~~~~~\nonumber\\
(d_1+d_2+d_3+d_4)+(d_2)+(d_3)+(d_4)~~~~~~\nonumber\\
(d_1+d_3+d_4+d_5)+(d_1+d_2+d_4+d_5)+\nonumber\\
(d_1+d_2+d_3+d_5)+(d_5)~~~~~~~~~~~~~~~~~~~~~~~\nonumber\\
(d_1+d_3+d_4+d_5)+(d_1+d_2+d_4+d_5)+\nonumber\\
(d_1+d_2+d_3+d_4)+(d_4)~~~~~~~~~~~~~~~~~~~~~~~\nonumber\\
(d_1+d_3+d_4+d_5)+(d_1+d_2+d_3+d_5)+\nonumber\\
(d_1+d_2+d_3+d_4)+(d_3)~~~~~~~~~~~~~~~~~~~~~~~\nonumber\\
(d_1+d_2+d_4+d_5)+(d_1+d_2+d_3+d_5)+\nonumber\\
(d_1+d_2+d_3+d_4)+(d_2)~~~~~~~~~~~~~~~~~~~~~~~
\end{eqnarray*}
In the both strategies $k-1$ nodes are involved during the course of downloading for creating a new node. As discussed, then this node accompanying two other nodes is used to reconstruct its related node. Therefore,  a repair bandwidth of size $\frac{(k-1)M}{k}+\frac{3M}{k}=\frac{(k+2)M}{k}$  is consumed to repair two failed node from a partition. In fact, the average repair bandwidth for reconstructing each node is $\frac{(k+2)M}{2k}$.

As an example, in $(10,5)$ code, two failed packets $d_1$ and $d_2+d_3+d_4+d_5$ are reconstructed after downloading from totally 7 nodes (i.e., 3.5 nodes for each packet), and consuming a repair bandwidth of size $\frac{7M}{5}$ (i.e., $\frac{3.5M}{5}$ for each packet). Recall in a (10,5) MSR code a new node is allowed to contact to at least five nodes which leads to a repair bandwidth of size $M$. It is worth mentioning that, for two specific cases $k=2$ and $k=3$ the reconstruction of a lost packet through communicating with $k-1$ nodes (strategy A and B) is more efficient than three nodes because in these cases $\frac{(k-1)M}{k}$ is smaller than $\frac{3M}{k}$.

Note the total storage for a file of size $M$ regardless of $k$ is $2M$ and we can reduce repair bandwidth having increase in $k$. Therefore, for a given total storage the suggested scheme can establish a tradeoff between the repair bandwidth and the number of storage nodes. Moreover, the number of nodes which are involved during the repair of a single node failure regardless of $k$ is three.

\section{Conclusion}\label{sec:conclusion}
This paper aims at introducing a non-MDS scheme which is applicable in storage systems. Our proposed code which entails $k$ partitions, each one consisting two related systematic and parity nodes, can tolerate any three node failures. Also it can tolerate any $k-1$ node failures if at most two of them being from a common partition. Moreover, each single node failure can be repaired through access to three nodes. The suggested code has a simplicity of implementation in such that each node stores only one packet and the recovery of the original data file and the reconstruction of a lost packet can be achieved by XORing the stored packets.
\bibliographystyle{IEEEtran}
\bibliography{IEEEabrv,refs}

\end{document}